\documentclass[amsmat,amssymb,amsfonts,aps,prb,twocolumn,showpacs]{revtex4}

\usepackage{graphicx}
\usepackage{dcolumn}
\usepackage{bm}

\begin{document}

\title{Interplay between sublattice and spin symmetry breaking in  graphene}

\author{ D. Soriano (1,3), J. Fern\'andez-Rossier (2,3)}
\affiliation{
(1) CIN2 (ICN-CSIC) and Universitat Autonoma de Barcelona, Catalan Institute of Nanotechnology, Campus UAB, 08193 Bellaterra (Barcelona), Spain
\\ (2) International Iberian Nanotechnology Laboratory, Av.   Mestre Jos\'e Veiga, 4715-330 
 Braga, Portugal\\
 (3) Departamento de F{\`i}sica Aplicada, Universidad de Alicante, San
Vicente del Raspeig, Spain }

\date{\today} 

\begin{abstract} 
We study the effect of sublattice symmetry breaking on the electronic, magnetic and transport properties of  two dimensional graphene as well as zigzag terminated one and zero dimensional  graphene nanostructures. The systems are described with the Hubbard model within the collinear mean field approximation.  We prove that for the non-interacting bipartite lattice with unequal number of atoms in each sublattice midgap states still exist in the presence of
 a staggered on-site potential $\pm \Delta/2$ .
We compute  the phase diagram of both 2D and 1D graphene with zigzag edges,   at half-filling,  defined by the normalized interaction strength $U/t$ and 
 $\Delta/t$, where $t$ is the first neighbor hopping.  In the case of 2D  we find that the system is always insulating and we find the  $U_c(\Delta)$ curve above which the system goes antiferromagnetic.
In 1D  we find that the system undergoes a phase transition from non-magnetic insulator for $U<U_c(\Delta)$ to a phase
with ferromagnetic edge order and antiferromagnetic inter-edge coupling. The conduction properties of the magnetic phase depend on $\Delta$ and can be insulating, conducting and even half-metallic, yet the total magnetic moment in the system is zero. 
We compute the transport properties of a heterojunction with two non-magnetic graphene ribbon electrodes connected to a finite length armchair ribbon and we find a strong spin filter effect.

\end{abstract}

\maketitle
\section{Introduction}

The most salient  electronic properties of graphene and its nanostructures are linked to the bipartite nature of the honeycomb lattice which is formed by two interpenetrating  identical triangular sublattices\cite{RMP07}. It is customary to refer to the sublattice as a pseudospin degree of freedom.  In this language, the
first neighbor hopping  is described in terms of a pseudo-spin flip operator, which results in the well studied electron-hole symmetric bands in graphene, whose wave-functions are sublattice unpolarized. The pseudospin symmetry becomes a chiral symmetry in the continuum limit in which electrons in graphene are described with a Dirac Hamiltonian\cite{Semenoff84} and  accounts for the lack of backscattering\cite{Ando98}, the  so called chiral tunneling\cite{Chiral-tunneling}  and the absence of an energy gap in two dimensional graphene. 

Sublattice symmetry breaking in graphene could arise spontaneously, due  to some electronic phase transition\cite{Min2008A,Araki2011,Semenoff2011}, or due to the coupling of graphene to some substrate, like Silicon Carbide\cite{Zhou2007,SIC-review} and Boron Nitride\cite{giovanetti2007,Hone2010,Xue2011}. Sublattice symmetry breaking would make it energetically favorable for the electrons to stay in one of the sublattices, resulting in pseudo-spin  order (either spontaneous, or induced).  The purpose of this work is to understand the interplay between induced pseudo-spin order and real spin order in graphene. Magnetic order is expected to take place in  monohidrogenated graphene zigzag edges. Within the standard one-orbital tight-binding model of graphene,  these edges give rise to a large density of states at the Fermi energy\cite{Nakada96} which is prone to a ferromagnetic inestability when Coulomb repulsion is considered within the mean field Hubbard model \cite{Fujita96}. Density functional calculations confirmed the scenario\cite{Son06,Cohen06}, showing that the  long-range Coulomb interactions and the other atomic orbitals, absent in the Hubbard model, do not play a major role in this system.  Both   the mean field  Hubbard model\cite{Fujita96,Waka03,Gunlycke07,JFR08,Fede09,Soriano-JFR10} and DFT calculations show that the magnetic phase with zero total spin has a gap, which opens due to inter-edge correlations\cite{JFR08}. The fabrication of graphene ribbons with ultrasmooth edges is now possible by unzipping carbon nanotubes\cite{Dai08,Dai10,Dai11}. Indirect evidence of magnetic order in the edges of zigzag ribbons is provided by Scanning Tunneling Spectroscopy (STS) that can be accounted for within the mean field Hubbard model \cite{Crommie2011}.

The sublattice degree of freedom plays a central role in the magnetic properties of bipartite lattices\cite{Lieb89,BreyRKKY,JFR07}.
Very much like an external magnetic  field  favors one spin orientation and splits the spin states, an external perturbation  favors one sublattice with respect to the other and   opens a gap in the band  structure of graphene\cite{Semenoff84,Niu2007}.  When this happens, it is not obvious a priori what happens to the edge states, even at the single particle level,  and the associated magnetism.   DFT calculations indicate that Boron Nitride zigzag ribbons with the edge atoms passivated with hydrogen are non-magnetic\cite{NakamuraPRB2005,Benzanilla2011} whereas graphene ribbons, 
 deposited on Boron Nitride ($BN$) whose lattice parameter is shifted to  match  graphene , indicate that edge magnetism survives \cite{Niu2011}.  These DFT calculations suggest that as  the sublattice symmetry breaking potential $\Delta$ increases, a phase transition must occur from magnetic to non-magnetic edges.  Here we address this problem using a much simpler description of the electron-electron interactions, namely, the mean field approximation for the Hubbard model, in the spirit of earlier work for graphene with the full sublattice symmetry and on recent work for graphene zigzag ribbons without inversion symmetry\cite{Pan2011}.  

The rest of this paper is organized as follows. In section II we present some general theorems regarding the  properties of the single particle states of the tight-binding model for graphene with a staggered potential. In section III we study the interacting model for the case of two dimensional graphene and study how the staggered potential affects the non-magnetic to antiferromagnetic transition. In section IV we study the interplay of magnetic and pseudo-spin order in the  case of zigzag ribbons.  We find that magnetic order and sub lattice symmetry breaking give rise to  spin polarization of the bands,  and in some instances we find half-metallic antiferromagnetic order.   The spin filter properties of this case are studied in section V, where we consider quantum transport between  two half-metallic zigzag ribbons separated by a non-magnetic armchair central region.   In section VI we summarize our main findings.

\section{Single particle states of a bipartite lattice with a staggered potential}
In this section we consider some quite general properties of the single particle states of the Hamiltonian of a bipartite sub-lattice with a constant sublattice symmetry breaking term: 
\begin{equation}
{\cal H}_0 =  \left( \begin{array}{cc} 0 & h_{AB} \\ h_{BA} & 0 \end{array}\right)+ \frac{\Delta}{2}
\left( \begin{array}{cc} {\bf 1} & 0 \\ 0 & -{\bf 1} \end{array}\right)= h_0 + V
\label{H0}
\end{equation}
where $h_{AB},h_{BA}$ and ${\bf 1}$ are matrices with dimension given by the number of atoms in sublattice $A$ and $B$.

\subsection{Null sublattice imbalance}

We consider first the case of a bipartite lattice {\em without sublattice imbalance}, so that the number of atoms in sublattice $A$ equals those in lattice $B$: $N_A=N_B$. In that case, $h_{AB}$, $h_{BA}$ and ${\bf 1}$ are all matrices of range $N_A=N_B$.   It can be easily seen that the sub-lattice symmetric hamiltonian, $h_0$ (or unperturbed hamiltonian) anti-conmutes with the sublattice imbalance operator  $\sigma_z$:
  \begin{eqnarray}
\left[h_0, \sigma_z\right]_{+}\equiv
\left[  \left( \begin{array}{cc} 0 & h_{AB} \\ h_{BA} & 0 \end{array}\right),
\left( \begin{array}{cc} 1 & 0 \\ 0 &-1 \end{array}\right)\right]_+=0
\end{eqnarray}
Since $\sigma_z^2=1$, it is said that the graphene Hamiltonian has a chiral symmetry.
As a result, if  $\vec{\psi}\equiv \left(\begin{array}{c}\vec{\psi_A} \\ \vec{\psi_B}\end{array}\right)$  is an eigenstate of $h_0$ with energy $E$,  we automatically have that  $ \vec{\phi}\equiv \sigma_z\vec{\psi}$ is also eigenstate with energy $-E$
\begin{equation}
h\vec{\psi}=E\vec{\psi},\,\,\,\,\rightarrow h\vec{\phi}=-E\vec{\phi}
\end{equation}
Thus, the chiral symmetry  ensures that the spectrum of $h_0$ has electron hole symmetry.
In addition, since $\vec{\psi}$ and $\vec{\phi}$ are eigenvectors of the same Hamiltonian with different eigenvalues they must be orthogonal. This leads to:
\begin{equation}
0= \vec{\psi}^*\cdot\vec{\phi}=\langle {\vec \psi} |\sigma_z |\vec{\psi}\rangle
\end{equation}
or more explicitly
 \begin{eqnarray}
0=\sum_{i\in A} |\psi_A(i)|^2-\sum_{j \in B}|\psi_B(j)|^2
\end{eqnarray}
Thus, this lead us directly that eigenstates of $h_0$ have equal total weight on the two sublattices. In a pseudospin language, they have a zero expectation value of the $\sigma_z$ pseudospin operator, since the Hamiltonian has the pseudomagnetic field (the hopping) in the
 $x,y$ plane.

We now turn our attention to    the eigenstates of a tight-binding hamiltonian with first neighbour hoppings defined in a bipartite lattice with a sublattice-dependent potential which is both homogeneous and traceless, as defined by equation (\ref{H0}). We are going to show that they also  have electron-hole symmetry. For that matter, we represent ${\cal H}_0$ in the  the subspace defined for a pair of eigenstates of $h_0$, $\vec{\psi}$ and $\vec{\phi}=\sigma_z\vec\psi$, with energies $E$ and $-E$ respectively.  We readily obtain
\begin{equation}
 H_0 =  \left( \begin{array}{cc} E &0 \\ 0 & -E \end{array}\right)+ \frac{\Delta}{2}
\left( \begin{array}{cc} 0  & 1 
\\ 1 &0 
 \end{array}\right)
\label{H22}
\end{equation}
whose eigenvalues are $\epsilon_{\pm}\equiv \pm \sqrt{E^2+\frac{\Delta^2}{4}}$ with corresponding eigenvectors $\vec{v}_{\pm}$ given by:

\begin{eqnarray}
\vec{v}_{+}=
Cos\frac{\theta}{2}\vec{\psi} +  Sin\frac{\theta}{2} \vec{\phi} 
\label{spinors1}
\end{eqnarray}
and
\begin{eqnarray}
\vec{v}_{-}=
Sin\frac{\theta}{2}\vec{\psi} -  Cos\frac{\theta}{2} \vec{\phi} 
\end{eqnarray}
where $Cos\theta=\frac{E}{\sqrt{E^2+\frac{\Delta^2}{4}}}$. Thus, if we know (half of) the spectrum and the eigenstate of  the sublattice symmetric problem $h_0$, we can easily build the spectrum and the eigenfunctions for the same lattice when a homogeneous traceless sublattice Zeeman term is added to the Hamiltonian. 

\begin{figure}
[t]
\includegraphics[width=0.90\linewidth,angle=0]{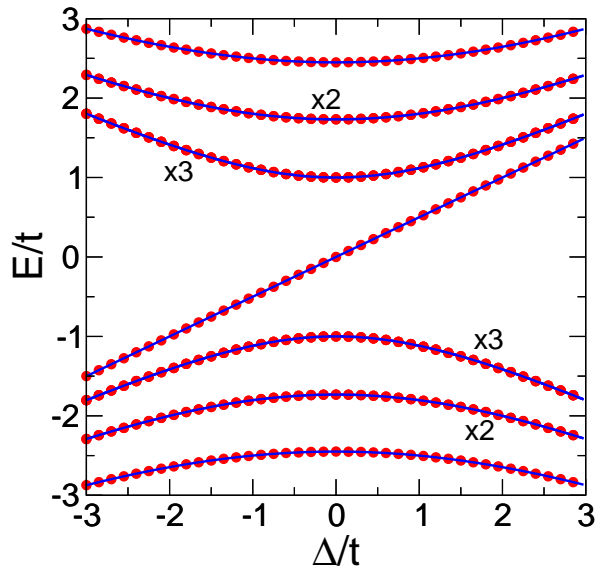}
\caption{ \label{figure1} (Color online).
 (a) Symbols: energy levels for triangulene with $N=13$ atoms calculated by diagonalization of the single-particle model. Lines: energy levels obtained from equation (\ref{levels}). The degeneracies are indicated in the figure.  }
\end{figure}

\subsection{System with sublattice imbalance} 
The results of the previous section need to be examined with care in the special case that $E=0$.  This certainly happens when 
we consider a system with $N_A=N_B+N_Z$, where $N_Z>0$ is a positive integer.
In that case the dimension of the $A$ and $B$ subspaces is not the same and the results of the previous section do not hold in general\cite{Pereira:2008}.  In particular, it has been shown that $h_0$ has $N_Z$ eigenstates $\vec{\psi}_Z$ with $E=0$ that are are sublattice polarized in the majority sublattice, $\vec{\psi}=\left(\begin{array}{c} \vec{z}_A \\ 0 \end{array}\right)$.  This are the so called midgap states and play a crucial role in the emergence of magnetism in graphene zigzag edges\cite{Fujita96} and graphene with chemisorbed hydrogen\cite{Yazyev:2007}.

 It can be inmediately seen that  if $\vec{\psi}_Z$ is a zero energy eigenstate of $h$ it is also an eigenstate of ${\cal H}_=h_0+V$ with eigenvalue $\epsilon=+\frac{\Delta}{2}$.  Conversely, if $h_0$ presents a  zero energy state sublattice polarized in $B$, then that state is also eigenvector of ${\cal H}_0$ with energy $-\frac{\Delta}{2}$. 

Thus,  we can now predict the evolution of the spectrum of a  given system described by ${\cal H}_0$ that,  at $\Delta=0$  has midgap states at zero energy as well as pairs of electron-hole symmetric states with finite energies $\pm E_n$.  The mid gap states will split with energy $\pm \Delta$, depending on their sublattice polarization, and the finite energy states will evolve as
\begin{equation}
\epsilon_{\pm}(\Delta)=\pm \sqrt{E_n^2+\frac{\Delta^2}{4}}. 
\label{levels}
\end{equation}
In order to illustrate this result, we have computed the single particle spectrum of a triangulene\cite{JFR07} with $N_A=7$ and $N_B=6$ atoms. The evolution of the spectrum as a function of $\Delta$ is shown in figure(\ref{figure1}). We compare the result of the numerical diagonalization with those extrapolated from the spectrum of $h_0$ and, expectedly, find perfect agreement. 

This calculation shows that, in structures with a larger number of midgap states, the midgap shell will remain half-full (when counting the spin), and interactions are expected to favor large spin configurations. 

\section{Electronic properties of two dimensional graphene with a staggered potential}
We now study the interplay between Coulomb repulsion and sublattice symmetry breaking. We model the interaction using a Hubbard model in the mean field approximation.  For symmetric graphene this approximation is known to predict a phase transition from the non-magnetic gapless state to an antiferromagnetic insulating state when\cite{Sorella92} $U>U_{c}= 2.2 t$.  As usual, the mean field approximation underestimates the critical $U$ necessary for the Mott transition. Quantum Monte Carlo calculations indicate\cite{Sorella92} that the transition takes place at $U_{c}\simeq 5.3 t$ .  In addition, recent work indicates that might be a third phase with spin-liquid properties  separating the non-magnetic state from the magnetically ordered phase \cite{Meng-Nature-2010}.  In spite of its limitations, the mean field description of the Hubbard model can shed some light on the possible ordered phases and their electronic properties.

\subsection{Hubbard model and a mean field approximation}
The extended  Hubbard model reads:
\begin{eqnarray}
\label{extended_hubbard}
H=t\sum_{ii',s}c^{\dagger}_{is}c_{i's} +\frac{\Delta}{2}\sum_{i,s} \tau_z(i) c^{\dagger}_{is}c_{is}+\nonumber\\
+U \sum_i n_{i\uparrow}n_{i\downarrow}= H_0 + U \sum_i n_{i\uparrow}n_{i\downarrow}
\end{eqnarray}
where $c^{\dagger}_{is}$ creates an electron in atomic site $i$ with spin $s=\uparrow,\downarrow$, $i'$ stand for the first neighbors of $i$, $\tau_z(i)=+1$ if $i$ belongs to the $A$ sublattice and $-1$ otherwise.  We only consider the half-filling case, where the number of electrons equals the number of sites in the lattice.  For a given filling, the ground state properties of the model depend on two dimensionless parameters $\Delta/t$ and $U/t$. We explore the properties for a spin-collinear mean field approximation, where the $U$ term is approximated by:  
\begin{equation}
V_{MF}=
+U \sum_i n_{i\uparrow}\langle n_{i\downarrow}\rangle+ \langle n_{i\uparrow}\rangle n_{i\downarrow}
\label{MF}
\end{equation}
where $ \langle n_{is}\rangle$ is the average of the occupation operator of site $i$ with spin $s$, calculated 
with the many-body ground state of the mean field Hamiltonian: 
\begin{equation}
\langle n_{is}\rangle = \sum_{\alpha} f_{\alpha}\langle \alpha | n_{is}|\alpha\rangle
\end{equation}
where $f_{\alpha}=0,1$ is the occupation of the single particle states $|\alpha\rangle$ that diagonalize 
 calculated with the ground state of the mean field Hamiltonian  $H_0 + V_{MF}$. Since the potential $V_{MF}$ depends on the eigenstates of $H+V_{MF}$, both the potential and the eigenstates need to be computed self-consistently.  We do this by iteration. 
 
 \subsection{Mean field approximation for 2D graphene}
 We now describe the electronic properties of Hubbard model for the two dimensional honeycomb lattice  with a staggered potential within the  mean field approximation.
    In this case, we can take a minimal unit cell with 2 atoms, $A$ and $B$ and assume that the mean field in all unit cells is identical, which permits to use Bloch theorem to represent the mean field Hamiltonian
  in the basis set  $A\uparrow, B\uparrow, A\downarrow, B\downarrow$:
\begin{eqnarray}
H=\left(
\begin{array}{cc} 
H_{\uparrow} & 0 \\
0 & H_{\downarrow}\end{array} \right)
\end{eqnarray}
where each element is a 2 by 2 matrix: 
  \begin{eqnarray}
H_{\uparrow}=\left(
\begin{array}{cc}
\frac{\Delta}{2} + U \langle n_{A\downarrow}\rangle & f(\vec{k})  \\
f^*(\vec{k}) & -\frac{\Delta}{2} + U \langle n_{B\downarrow}\rangle 
\end{array}
\right)
\end{eqnarray}
and
\begin{eqnarray}
H_{\downarrow}=\left(
\begin{array}{cc}
 \frac{\Delta}{2} + U \langle n_{A\uparrow}\rangle & f(\vec{k})   \\
f^*(\vec{k}) & -\frac{\Delta}{2} + U \langle n_{B\uparrow}\rangle
\end{array}
\right)
\end{eqnarray}
and
\begin{equation}
f(k)= t\left(1+ e^{i\vec{k}\cdot\vec{a}_1} +e^{i\vec{k}\cdot\vec{a}_2} \right)
 \end{equation}
 accounts for the first neighbour particle hopping. 
 In our numerical determination of the self consistent occupations $\langle n_{is}\rangle$ we have taken an unit cell of 4 atoms.  We have verified that our mean field solutions in this extended unit cell do not present inter-cell modulations of the charge density.  
 
\begin{figure}
[t]
\includegraphics[width=0.90\linewidth,angle=0]{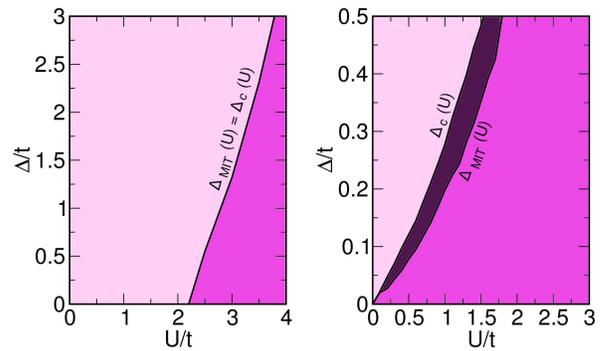}
\caption{ \label{PHASE2D} (Color online).
Phase diagrams for 2D-graphene (left) and a zigzag graphene nanoribbon with $N=48$ atoms in the unit cell (right) with stagger potential ($\Delta$) using a mean field Hubbard model at half filling. The dark region with $\Delta_{MIT}(U)> \Delta(U) > \Delta_c(U)$ correspond to the spin half-metallic phase in the graphene ribbon. In the case of 2D-graphene, this region is reduced into a single critical line separating non-magnetic and antiferromagnetic insulating states.}
\end{figure}

\begin{figure}
[t]
\includegraphics[width=0.90\linewidth,angle=0]{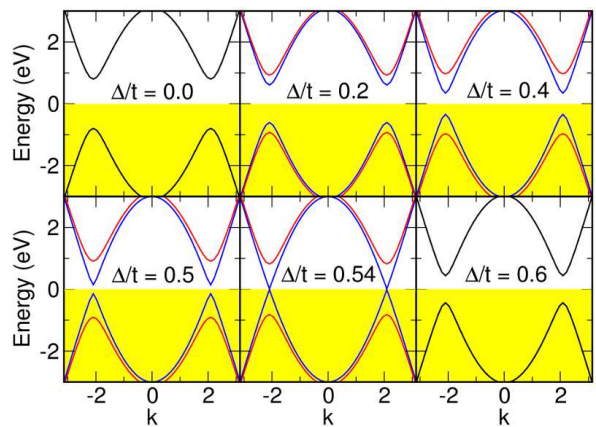}
\caption{ \label{BANDS2D} (Color online).
Band structure of 2D-graphene with stagger potential ($\Delta$) in the mean field extended Hubbard approximation for $U > U_c (\Delta)$}
\end{figure}

 We have explored the phase diagram defined by $U/t$ and $\Delta/t$ and we  find 3 types of solution, shown in figure (\ref{PHASE2D}): 
 \begin{enumerate}
 \item For $U<U_c(\Delta)$  the system is non magnetic and, except for $\Delta=0$, a band insulator. The case of $\Delta=0$ and $U<U_c$ is the well studied paramagnetic semimetal phase. 
 
 \item For $U>U_c(\Delta)$ the system is an antiferromagnetic insulator.  The lack of inversion symmetry produced by the sublattice symmetry breaking results in a splitting of the spin bands, in contrast with the standard $\Delta=0$ case.  This is shown in figure (\ref{BANDS2D}) 

\item  For $U=U_c(\Delta)$ the system is  a half-semimetallic antiferromagnet.  For one spin channel the system is insulating and for the other is semimetallic. 

 \end{enumerate}

It is apparent that, as $\Delta$ increases, the critical $U_c$ increases. Expectedly, the magnetic order has to overcome the single-particle gap opened by the staggered potential. Interestingly, the mean field approximation describe a magnetic transition between two insulating states, the non-magnetic insulator and the antiferromagnetic insulating phase, which can be interpreted as an excitonic insulator transition.  Given the large values of $U_c(\Delta)/t$ this ordered electronic phase is not expected in graphene. The predictions of this theory should be tested in cold atomic gases confined in optical lattices\cite{optical-lattice}  or in artificially paterned honeycomb lattices in two dimensional electron gases\cite{2DEG}

\section{Electronic properties of graphene zigzag ribbons  with a staggered potential}

We now  study the case of zigzag graphene ribbons for which we find that magnetic order could happen at low values of $U/t$ even for finite $\Delta$. 
 The width  of the ribbons is characterized by $N$, the number of atoms in the  unit cell.  Importantly, one of the edges is formed with $A$ atoms only, the other being made of $B$ atoms only. Thus, pseudospin polarization implies charge accumulation in one edge and depletion in the other, ie, the formation of an electric dipole.

 The electronic structure of graphene ribbons has been widely studied in the $\Delta=0$ limit, both for the $U=0$\cite{Nakada96,Brey-Fertig} and the finite $U$ cases
\cite{Fujita96,Waka03,Son06,Cohen06,Gunlycke07,JFR08,Yaz07,Fede09,Soriano-JFR10}. The most prominent feature of their electronic structure is given by the flat bands associated to edge states. At $\Delta=U=0$, these bands are located at the Fermi energy, giving rise to a large density of states at the Fermi energy.  Not surprisingly, Coulomb repulsion results in a magnetic inestability\cite{Fujita96} corresponding to the formation of magnetic moments in both edges while the bulk-atoms remain  almost spin unpolarized.  It turns out that the inter-edge spin correlations are antiferromagnetic, as expected from the Lieb theorem\cite{Lieb89}. Thus, for a given spin orientation, there is charge accumulation in one of the edges and charge depletion in the opposite. 
This results in a spin-resolved pseudo-spin polarization, or spin-dipole \cite{JFR08}.
Here we are interested in the interplay between pseudo-spin polarization, driven by the $\Delta$ term in the Hamiltonian, and the spin polarization, which entails a spin-resolved pseudo-spin polarization, driven by the Coulomb repulsion $U$.

\subsection{Non interacting bands}
\begin{figure}
[t]
\includegraphics[width=0.90\linewidth,angle=0]{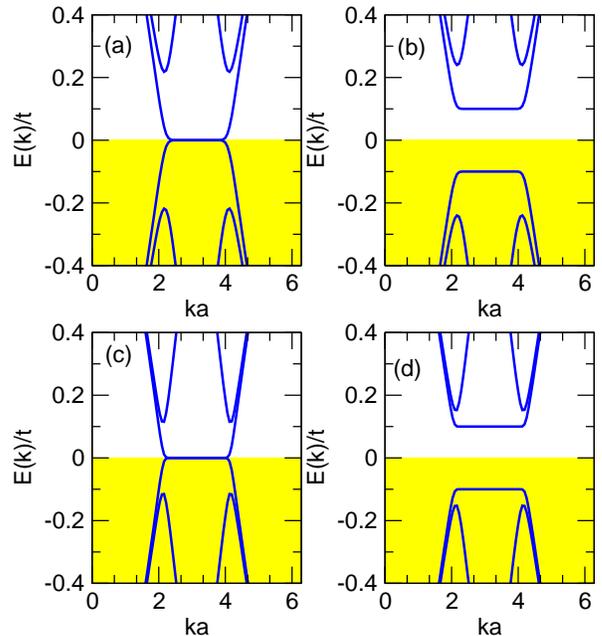}
\caption{ \label{freebands} (Color online).
Electronic structure of zigzag ribbons with $U=0$ , for   $\Delta=0$ and $\Delta=0.2t$ (left and right panels) for 2 different ribbon widths,  $N=40$ (top) and, $N=80$ (bottom). For the sake of clarity, we only plot the 2 higher energy valence bands and the 2 lowest energy conduction bands.  }
\end{figure}

  We first review the effect of the staggered potential  on the non-interacting bands, studied by Qiao {\em et al.}\cite{Niu1}.  At $\Delta=0$ two almost flat bands, associated to edge states, lie at the Fermi energy.  As  $\Delta$ becomes finite (and positive), the bands at $B$ edge are red-shifted and those at $B$ edge are blue-shifted, resulting in a band-gap opening.  This is seen in figure(\ref{freebands}) for two ribbons with $N=40$ and $N=80$ atoms, for $\Delta=0$ (left columns) and $\Delta=0.2t$ (right column).  We also notice the low energy bands are quite similar for the $N=40$ and $N=80$ ribbons, whereas the gap between the
higher energy bands is reduced for the wider ribbon. This is consistent with the fact that lowest energy bands are edge states, relatively insensitive to the  width of the ribbon, in contrast with  higher energy bands made of 
 quantum confined bulk states\cite{Brey-Fertig}. 
 
 Thus, for finite $\Delta$ and $U=0$ graphene zigzag ribbons are band insulators with pseudospin polarization  that features two  flat bands corresponding to  the highest occupied and lowest un-occupied bands.  For  $\Delta>0$, the  bands corresponding to both $\uparrow$ and $\downarrow$ spins in  the $B$ edge are occupied, whereas those in edge  $A$ are empty. As we show now, these bands are prone to magnetic inestability, not-unlike in the case with $\Delta=0$. 

\subsection{Effect of Coulomb repulsion}
\begin{figure}
[hbt]
\includegraphics[width=0.90\linewidth,angle=0]{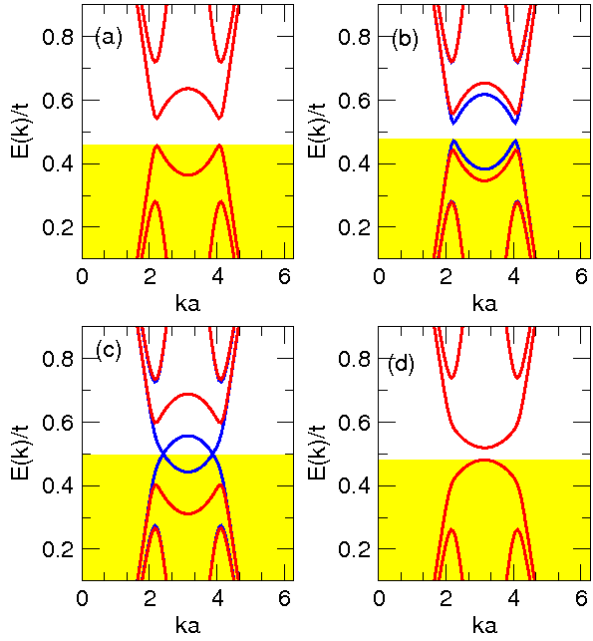}
\caption{ \label{u10bands} (Color online).
 Lowest energy bands for  for zigzag ribbon with $N=48$ atoms in unit cell and $U= t$, for different values of $\Delta$: (a) $\Delta=0$, (b) $\Delta=0.05t$,  (c) $\Delta=0.2t$ (d) $\Delta=0.3 t$. Only the 2 highest energy occupied bands and the 2 lowest energy empty bands, per spin channel, are shown. Blue (red) stands for $\uparrow$ ($\downarrow$) bands. }
\end{figure}

We study now the interplay between pseudospin polarization and Coulomb repulsion. For that matter, 
 we use again the mean field approximation for the Hubbard model, as described in previous work\cite{Fujita96,Waka03,JFR08,Fede09,Soriano-JFR10,Pan2011}. 
 Numerically found solutions present  magnetization at both edges  equal in magnitude and opposite in sign. Thus, there are two equivalent ground states: $m_A=-m_B>0$ and $m_A=-m_B<0$.  
 The corresponding energy bands for the ribbon with $N=48$, and $\Delta=0$ and $U=t$ are shown in figure (\ref{u10bands}a). The magnetic order results in a band-gap opening. The spin $\uparrow$ and $\downarrow$ bands are degenerate.   The magnetic moment at the edge atoms is $m=\pm 0.13$.  The charge per atom is the same all over the unit cell, 1 electron per atom. 
    
When the sublattice-symmetry breaking potential is finite and below a critical value $\Delta_c(U)$,  we still find magnetic order in the edges with antiferromagnetic coupling, and zero total moment, even if the charge is no longer the same for both edges. 
The electronic properties of the magnetic ribbon with pseudo-spin polarization are different on several counts.   First, 
the bands are spin split, as a  natural consequence of the lack of both time-reversal and inversion symmetries. 
The evolution of the energy bands,  as we increase $\Delta$, is shown in figure (\ref{u10bands}) for the $N=48$ ribbon with  $U=t$. This figure can be understood as follows.
The solution has  $m_A=-m_B>0$ so  that the $\uparrow$ band is occupied (empty) in the $A$ ($B$)  edge. Conversely, the $\downarrow$ band is occupied (empty) in the $B$ ($A$)  edge. 
 As $\Delta$ is turned on, the $B$ bands are red-shifted and $A$ bands are blue shifted. For the $\uparrow$ bands, this implies that the band gap closes, since the valence $A$ bands move upwards and the conduction $B$ bands move downwards. Conversely, the gap opens in the $\downarrow$ channel. 
 
As shown in figure (\ref{properties})  and discussed below, as $\Delta$ increases the magnetic moment at the edges are depleted and, eventually disappear when  $\Delta>\Delta_c(U)$.   Remarkably, the gap in the $\uparrow$ channel closes for $\Delta_{\rm MIT}(U)<\Delta_c(U)$,  yet the gap is finite in the $\downarrow$ channel. Thus,  the combination of pseudo-spin polarization and antiferromagnetic order makes the system a half-metallic anti-ferromagnet in a region of the $\Delta,U$ phase space (see dark-violet region in figure (\ref{PHASE2D})right panel). Notice that this is different from the ferromagnetic  half-metallic  phase predicted for graphene ribbons in the presence of a transverse electric field \cite{Cohen06}, for which the total magnetic moment is different from zero.

Finally, we note that the gap of the non-magnetic case in figure (\ref{u10bands})d for  $\Delta>\Delta_c$ is significantly smaller than $\Delta$. This  is due to the renormalization of the bands due to Coulomb repulsion. Basically, the occupied bands are blue shifted with respect to the empty bands, reducing the size of the gap.

\subsection{Phase Diagram}

In figure (\ref{PHASE2D}) we show the phase diagram defined by $U/t$ and $\Delta/t$  for a ribbon with $N=48$ atoms, calculated within the mean field approximation at half-filling. Earlier work\cite{Pan2011} has addressed the phase diagram defined by $\Delta$ and the electron density. 
The diagram in figure (\ref{PHASE2D}) has  2 phases  regarding the magnetic order: non-magnetic, for $\Delta>\Delta_c(U)$ and antiferromagnetic otherwise.   The results are very similar for ribbons with different widths. 
     In contrast with the 2D case, for $\Delta=0$  the critical $U$ for the edge is zero. This makes zigzag graphene ribbons suitable systems for the observation of magnetism in graphene and the possible effect of sublattice symmetry breaking more relevant. 
Expectedly, the critical $\Delta_c(U)$ is an increasing function of $U$, or in other words: the larger the single particle gap, the strongest the interaction $U$ required to drive the magnetic inestability.
   Spin polarization requires promotion of electrons across the single particle gap, from the occupied $B$  to the empty $A$ edge,  ie, the formation of a magnetic exciton condensate. 
     The difference with the $\Delta=0$ case stands on the size of the  the single particle gap, which  is vanishingly small (but not zero) in finite width ribbons\cite{JFR08}.  Interestingly, the magnetic exciton condensation scenario already takes place  in the case  apparently conducting $\Delta=U=0$ ribbon, when $U$ is turned on.

\begin{figure}
[t]
\includegraphics[width=0.90\linewidth,angle=0]{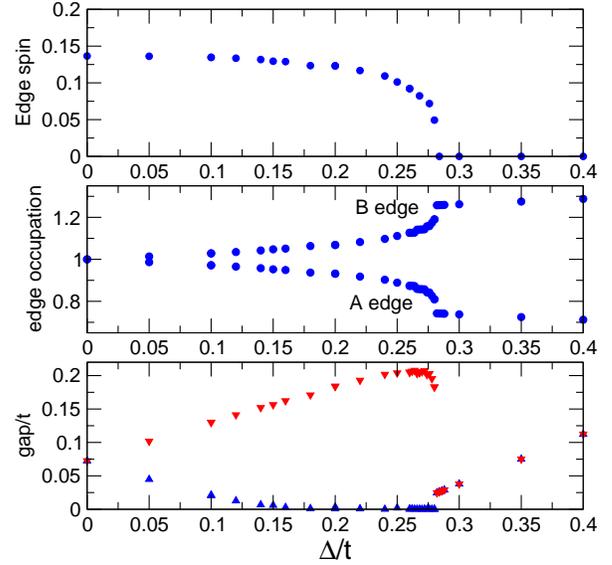}
\caption{ \label{properties} (Color online).
Properties of $N=48$ ribbon with $U=t$ as a function of $\Delta$. Top panel: edge magnetization. Middle panel: edge charges. Bottom panel: gap for spin $\uparrow$ and $\downarrow$.
 }
\end{figure}

For a fixed value of $\Delta$,  as the value of $U$ is increased the system undergoes a phase transition from a non-magnetic band insulator, to a half-metallic antiferromagnet and then to an insulating antiferromagnet.  The fact that interactions can drive the system from insulating from metallic is quite exotic and differs from the usual Mott insulator scenario, in which interactions drive a band metal insulating.
In the phase diagram we mark the insulator to metal transition when the smallest energy gap, at a given spin channel, is 100  times smaller than the gap at $\Delta=0$.  As shown in figure (\ref{properties}) for $U=t$, the gap at $\Delta=0$ is $0.07t$. Thus,  for that particular value of $U$,  we  declare the system conducting when the gap is 7 $10^{-4}$t.  Variations upon this criteria yield quantitative changes in the metal to insulator transition line, but it is always the case that $\Delta_{MIT}(U)$ runs  along and   below the magnetic phase transition line $\Delta_c(U)$.


\section{Graphene zigzag ribbons with staggered potential as an ideal spin injector}
Interestingly, the predicted conducting  phase for $\Delta_{\rm MIT}(U)<\Delta<\Delta_c(U)$   is  a half-metallic antiferromagnet. In this section we study the spin transport properties of a  tunnel junction where the electrodes are made of such half-metallic antiferromagnets and the barrier is made of semiconducting armchair graphene ribbon. 

\begin{figure}
\includegraphics[width=0.80\linewidth,angle=0]{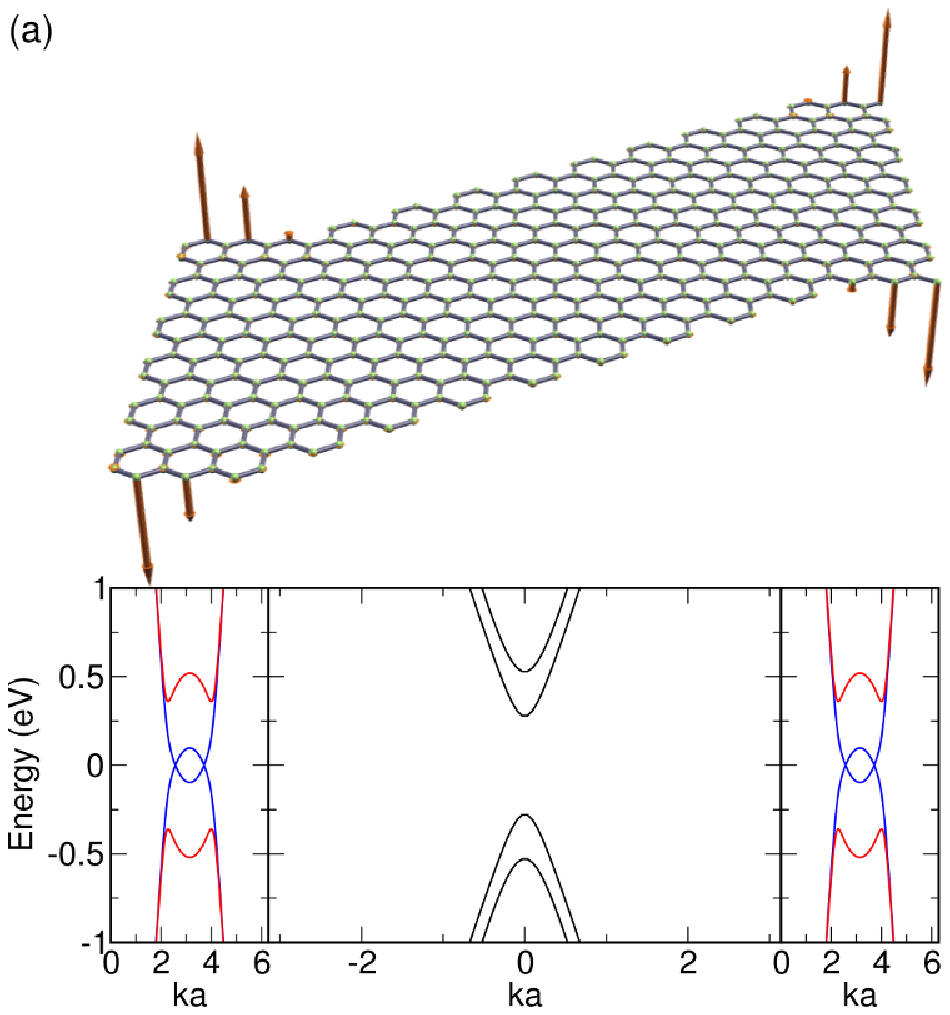}
\includegraphics[width=0.80\linewidth,angle=0]{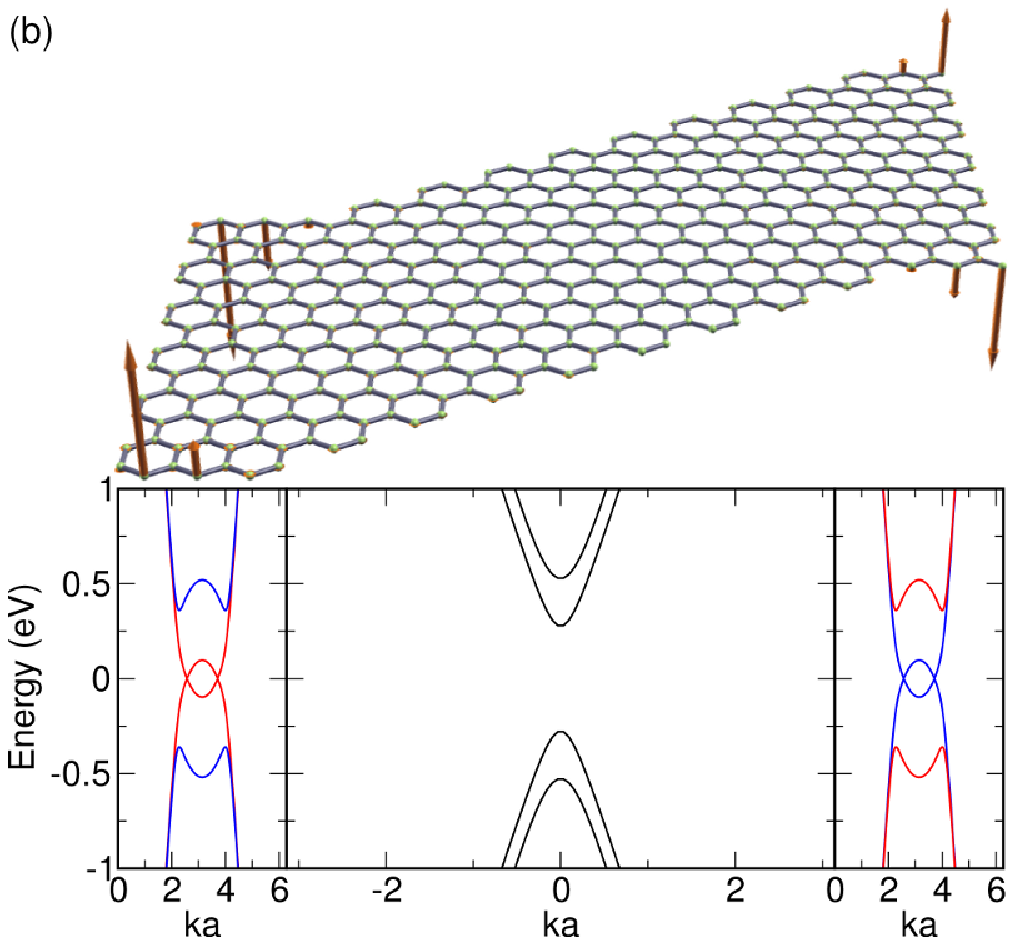}
\includegraphics[width=0.80\linewidth,angle=0]{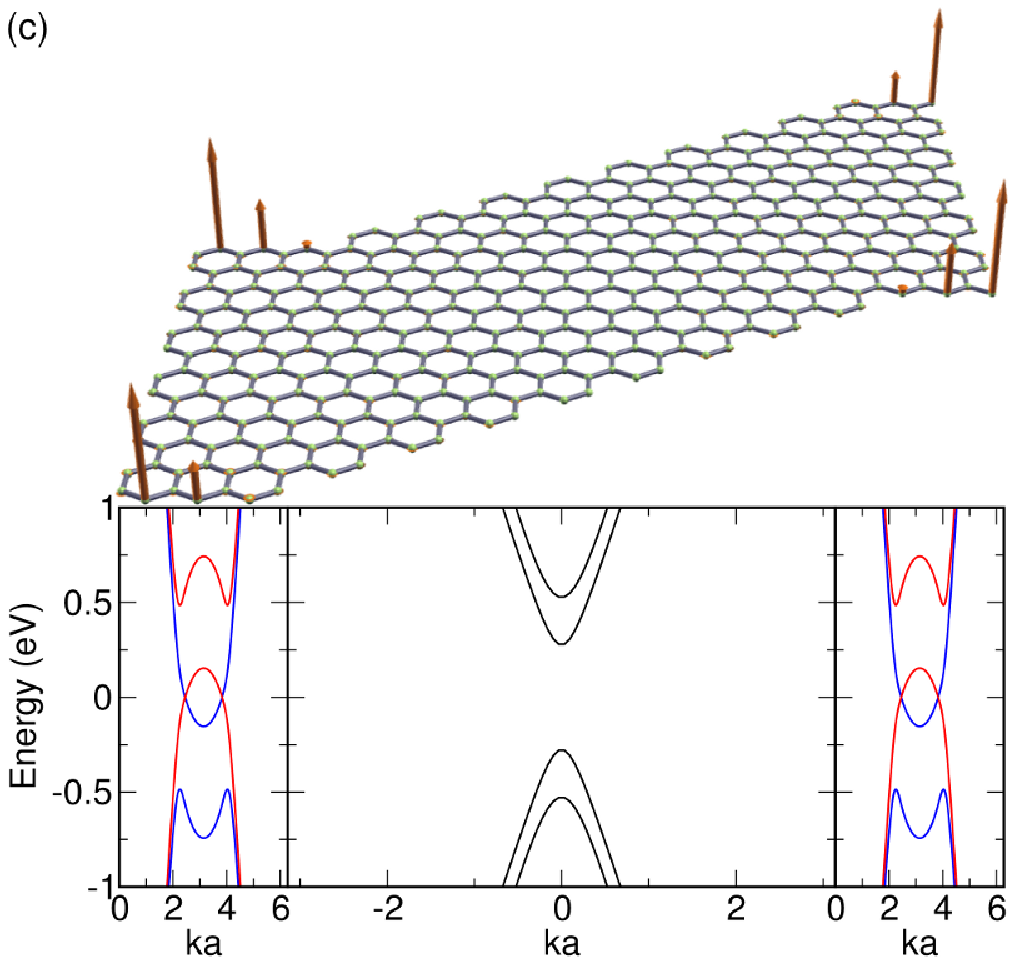}
\caption{ \label{bands-device} (Color online) Band structure of the leads and central region for the three cases studied: (a) parallel antiferromagnetic electrodes, (b) antiparallel antiferromagnetic electrodes and (c) parallel ferromagnetic electrodes,. The central region is the same for the three cases studied, namely, a semiconducting armchair graphene nanoribbon with a local Coulomb potential ($U=t$).   
 }
\end{figure}

In figure(\ref{bands-device}) we consider three possible situations, all of them with  {\bf $\Delta=0.25t$} , $U=t$.  The first and second cases feature two  antiferromagnetic electrodes with mutually  parallel and antiparallel magnetizations, respectively.  Based on the band structures of these infinite ribbons, shown in figure, we expect the tunnel conductance to be completely depleted when the magnetic moments of the different electrodes are anti-parallel. In the last case we consider two ferromagnetic electrodes. The bands of the infinite ribbon with ferromagnetic coupling between the magnetic edges reflect the conducting and spin unpolarized character of the system at the Fermi energy (set at E=0 eV in the three cases studied).


The conductance of the system is calculated within the  Landauer formalism and the Green's funtion method in a system consisting of an armchair semiconducting graphene nanoribbon with an on-site Coulomb potential connected to two spin-polarized zigzag nanoribbons with both stagger and on-site Coulomb potentials, as those studied in the previous section.  Except for the atoms at the interface with the zigzag ribbons, the Hubbard $U$ term is not able to spin-polarized the amrchair central region, as shown in the figure  (\ref{bands-device}). The short length of the tunneling barrier justifies also the neglect of spin relaxation which are expected in longer samples.  

To compute the transmission function ${T}(E)$ along the armchair nanoribbon we adopt a partitioning method as implemented in the ALACANT(Ant.U)\cite{ALACANT} transport package,  where the  system is divided into three parts\cite{Fede09,Soriano10}, namely, the central ($C$) part which consists on the armchair ribbon connected to a small part of the electrodes, and the left and right semi-infinite staggered zigzag nanoribbons ($L$ and $R$).    

The transmission probability can then be obtained from the Caroli expression\cite{Caroli:71} 
\begin{equation}
\label{Caroli}
T_\sigma(E) = \mathrm{Tr}[\mathcal{G}_C^\dagger(E) \Gamma_R(E) \mathcal{G}_C(E) \Gamma_L(E)]_\sigma
\end{equation}
where $\mathcal{G}_{C,\sigma}(E)=[z{I}-H_{C,\sigma}-\Sigma_{R,\sigma}(E)-\Sigma_{L,\sigma}(E)]^{-1}$ is the Green's function of the central region which contains all the information concerning the electronic structure of the semi-infinite leads through the self-energies ($\Sigma_{R,\sigma}(E)$ and $\Sigma_{L,\sigma}(E)$), and $\Gamma_{R,\sigma}(E)=i[\Sigma_{L,\sigma}(E)-\Sigma^\dagger_{L,\sigma}(E)]$, $\Gamma_{L,\sigma}(E)=i[\Sigma_{R,\sigma}(E)-\Sigma^\dagger_{R,\sigma}(E)]$ are the coupling matrices containing the information about the coupling of the central region to the leads. 

\begin{figure}
[t]
\includegraphics[width=0.90\linewidth,angle=0]{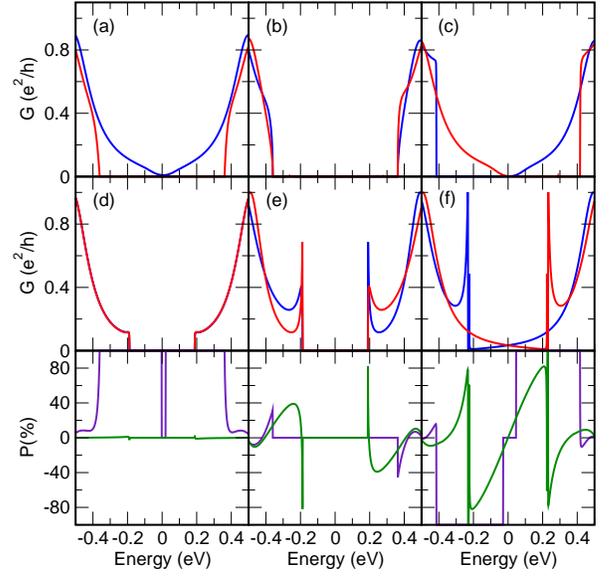}
\caption{ \label{transpol} (Color online) Spin conductance through a finite armchair ribbon connected to two spin-polarized zigzag ribbons with (a-c) and without (d-f) stagger potential. The bottom panel shows the spin conductance polarization for each magnetic ordering at the electrodes in the presence (violet) and absence (green) of stagger potential.
 }
\end{figure}

In Fig. \ref{transpol}(a-f), we have computed the spin-polarized conductance for the three systems shown in Fig. \ref{bands-device}. The top and middle panels correspond to the cases with and without stagger potential respectively. The bottom panel of Fig. \ref{transpol} shows the conductance polarization $P = G_\uparrow - G_\downarrow / G_\uparrow + G_\downarrow \times 100$ at different energies with (violet line) and without (green line) stagger potential.

The results obtained for the spin conductance can be easily inferred by looking at the band structures of the three regions in Fig. \ref{bands-device}. The two antiferromagnetic cases without stagger show a large gap in the conductance due to the semiconducting behavior of the three regions. For the ferromagnetic case, there is a finite conductance near the Fermi energy due to the metallic nature of the electrodes. In this case, the evanescent modes coming from both electrodes penetrate into the central region and overlap due to the short length of the armchair ribbon.

When both electrodes are antiferromagnetic with mutually parallel magnetization, the system transforms into a spin-half metal where both valence and conduction bands show the same spin polarization. In this case, the polarization of the conductance is the same at both sides of the Fermi energy. The case featuring antiferromagnetic leads with mutually antiparallel magnetization also shows a spin half-metallic phase in both leads but with opposite spin polarization around the Fermi level. Thus the spin polarized  current injected from one electrode is always reflected by the opposite electrode resulting in a zero conductance around the Fermi energy. If both electrodes are coupled ferromagnetically the system becomes metallic and the valence and conduction bands cross at the Fermi level but with a different spin polarization. This makes the conductance polarization to change sign at each side of the Fermi level.

\section{Summary and conclusions}



Our main findings can be summarized as follows:
\begin{enumerate}
\item We have shown that for a bipartite lattice with a site-independent  pseudo-spin Zeeman term the resulting spectrum has electron hole symmetry, except for mid-gap states that arise in lattices with a different number of sites in the two sublattices.  We have shown that, if the $\Delta=0$ Hamiltonian has an eigenstate $\vec{\phi} $ with energy $E$, it has also an eigenstate $\vec{\psi}$  with energy $-E$ and the Hamiltonian with finite $\Delta$ has two  eigenstates, linear combination of $\vec{\phi}$ and $\vec{\psi}$ with energies $\pm \sqrt{\frac{\Delta^2}{4}+E^2}$

\item If $\vec{\phi}$ localized in the $A$ ($B$) sublattice is an eigenstate with energy $E=0$ of  the $\Delta=0$ Hamiltonian, then it is also an eigenstate of the  Hamiltonian with finite $\Delta$ and energy $\Delta$ ($-\Delta$)

\item Two dimensional graphene with finite $\Delta$ undergoes a transition to an antiferromagnetic state with spin-split bands, for $U>U_c(\Delta)$.  The critical $U_c$ is an increasing function of $\Delta$. At the transition between the non-magnetic insulating state ($U<U_c(\Delta)$) and the magnetic state $U>U_c$, the system is a half metallic antiferromagnet. 

\item Zigzag graphene ribbons with a finite sublattice symmetry-breaking potential below a critical value ($\Delta(U) < \Delta_c(U)$) can undergo a transition from non-magnetic insulators to spin-polarized half-metallic antiferromagnets in the presence of Coulomb interaction. The fact that interactions can drive the system from insulating to metallic differs from the usual Mott insulator scenario in which interactions drive a band metal insulating.    

\item Zigzag graphene ribbons with stagger potential $\Delta$ slightly  below the $\Delta_c(U)$  are predicted to be ideal spin injectors. The spin transport calculations carried out in this work show high spin polarization of the conductance ($\approx 100\%$) around the Fermi level when used as spin injectors in a tunnel junction.   This indicates that, if the suitable substrate that yields the right $\Delta$ is found, graphene zigzag ribbons could act as half-metallic spin injectors. 	

\end{enumerate}


This work has been financially supported by MEC-Spain (FIS-
 and CONSOLIDER CSD2007-0010).  We acknowledge fruitful conversations with Jeil Jung.

\end{document}